\documentclass[final,5p,times,twocolumn]{elsarticle}

	\usepackage{color}
	\usepackage{arydshln}
	\usepackage{lineno}
    
    \journal{}

\begin{document}
	
	\begin{frontmatter}
	
	\title{Identification of $^{210}$Pb and  $^{210}$Po in the bulk of copper samples with a low-background alpha particle counter}

	\address{\rm\normalsize XMASS Collaboration$^*$}
	\cortext[cor1]{{\it E-mail address:} xmass.publications5@km.icrr.u-tokyo.ac.jp .} 
	
	\author[ICRR,IPMU]{K.~Abe}
	\author[ICRR,IPMU]{K.~Hiraide}
	\author[ICRR,IPMU]{K.~Ichimura}
	\author[ICRR,IPMU]{Y.~Kishimoto}
	\author[ICRR,IPMU]{K.~Kobayashi}
	\author[ICRR]{M.~Kobayashi}
	\author[ICRR,IPMU]{S.~Moriyama}
	\author[ICRR,IPMU]{M.~Nakahata}
	\author[ICRR]{T.~Norita}
	\author[ICRR,IPMU]{H.~Ogawa}
	\author[ICRR]{K.~Sato}
	\author[ICRR,IPMU]{H.~Sekiya}
	\author[ICRR]{O.~Takachio}
	\author[ICRR,IPMU]{A.~Takeda}
	\author[ICRR]{S.~Tasaka}
	\author[ICRR,IPMU]{M.~Yamashita}
	\author[ICRR,IPMU]{B.~S.~Yang}
	\author[IBS]{N.~Y.~Kim}
	\author[IBS]{Y.~D.~Kim}
	\author[ISEE,KMI]{Y.~Itow}
	\author[ISEE]{K.~Kanzawa}
	\author[ISEE]{R.~Kegasa}
	\author[ISEE]{K.~Masuda}
	\author[ISEE]{H.~Takiya}
	\author[Tokushima]{K.~Fushimi\fnref{Tokushimanow}}
	\author[Tokushima]{G.~Kanzaki}
	\author[IPMU]{K.~Martens}
	\author[IPMU]{Y.~Suzuki}
	\author[IPMU]{B.~D.~Xu}
	\author[Kobe]{R.~Fujita}
	\author[Kobe]{K.~Hosokawa\fnref{Tohokunow}}
	\author[Kobe]{K.~Miuchi}
	\author[Kobe]{N.~Oka}
	\author[Kobe,IPMU]{Y.~Takeuchi}
	\author[KRISS,IBS]{Y.~H.~Kim}
	\author[KRISS]{K.~B.~Lee}
	\author[KRISS]{M.~K.~Lee}
	\author[Miyagi]{Y.~Fukuda}
	\author[Tokai1]{M.~Miyasaka}
	\author[Tokai1]{K.~Nishijima}
	\author[YNU1]{S.~Nakamura}

	\address[ICRR]{Kamioka Observatory, Institute for Cosmic Ray Research, the University of Tokyo, Higashi-Mozumi, Kamioka, Hida, Gifu, 506-1205, Japan}
	\address[IBS]{Center of Underground Physics, Institute for Basic Science, 70 Yuseong-daero 1689-gil, Yuseong-gu, Daejeon, 305-811, South Korea}
	\address[ISEE]{Institute for Space-Earth Environmental Research, Nagoya University, Nagoya, Aichi 464-8601, Japan}
	\address[Tokushima]{Institute of Socio-Arts and Sciences, The University of Tokushima, 1-1 Minamijosanjimacho Tokushima city, Tokushima, 770-8502, Japan}
	\address[IPMU]{Kavli Institute for the Physics and Mathematics of the Universe (WPI), the University of Tokyo, Kashiwa, Chiba, 277-8582, Japan}
	\address[KMI]{Kobayashi-Maskawa Institute for the Origin of Particles and the Universe, Nagoya University, Furo-cho, Chikusa-ku, Nagoya, Aichi, 464-8602, Japan}
	\address[Kobe]{Department of Physics, Kobe University, Kobe, Hyogo 657-8501, Japan}
	\address[KRISS]{Korea Research Institute of Standards and Science, Daejeon 305-340, South Korea}
	\address[Miyagi]{Department of Physics, Miyagi University of Education, Sendai, Miyagi 980-0845, Japan}
	\address[Tokai1]{Department of Physics, Tokai University, Hiratsuka, Kanagawa 259-1292, Japan}
	\address[YNU1]{Department of Physics, Faculty of Engineering, Yokohama National University, Yokohama, Kanagawa 240-8501, Japan}

	\fntext[Tokushimanow]{Now at Department of Physics, Tokushima University, 2-1 Minami Josanjimacho Tokushima city, Tokushima, 770-8506, Japan}
	\fntext[Tohokunow]{Now at Research Center for Neutrino Science, Tohoku University, Sendai, Miyagi 980-8578, Japan.}

	\begin{abstract}
We established a method to assay $^{210}$Pb and $^{210}$Po contaminations
in the bulk of copper samples using a low-background alpha particle counter.
The achieved sensitivity for the $^{210}$Pb and $^{210}$Po contaminations
reaches a few mBq/kg.
Due to this high sensitivity, the $^{210}$Pb and $^{210}$Po
contaminations in oxygen free copper bulk were identified and
measured for the first time.
The $^{210}$Pb contaminations of our oxygen free copper
samples were 17-40 mBq/kg.
Based on our investigation of copper samples in each production step,
the $^{210}$Pb in oxygen free copper was understood to be a small residual
of an electrolysis process.
This method to measure bulk contaminations of $^{210}$Pb and $^{210}$Po
could be applied to other materials.

    \end{abstract}
	
	\begin{keyword}
		$^{210}$Pb \sep $^{210}$Po \sep copper \sep radio purity \sep Alpha spectroscopy
	\end{keyword}

	\end{frontmatter}

    
	\section{Introduction}
    \label{Sec:introduction}
In recent low-background experiments performed underground and aiming
for dark matter or neutrinoless double beta decay detection, radio-purity
of the detector materials is a critical issue. 
Oxygen free copper (OFC) is readily available commercial 
material of low radio isotope (RI) content.
Thus it is used in experiments
such as XMASS \cite{bibXMASSDetector}, and CUORE \cite{bibCUOREDetector}.
In order to investigate RI in the uranium and thorium chains, 
$^{238}$U and $^{232}$Th contaminations are usually measured by Inductively 
Coupled Plasma Mass Spectrometry or Glow Discharge Mass 
Spectrometry (GD-MS) with high precision. 
However, if the radioactive equilibrium is broken, inferred $^{226}$Ra
or $^{210}$Pb contaminations will be different.
With a high purity germanium detector (HPGe), $^{226}$Ra can be measured 
down to a few tens of mBq/kg.
However, the sensitivity for $^{210}$Pb contamination is limited to
around 100 mBq/kg. 
Moreover, HPGe cannot distinguish surface from bulk contamination.
Since the requirements for RI contamination are becoming
more stringent, we developed a better way to measure the contamination
of $^{210}$Pb in copper with a low-background alpha particle counter.
By measuring the time evolution of event rate of the 5.30 MeV alpha rays
from $^{210}$Po decay with an alpha particle counter, the $^{210}$Pb
contamination can be determined because the $^{210}$Po half-life is
138.4 days and much shorter than that of the parent isotope, $^{210}$Pb.

In this paper, we report that  we measured $^{210}$Pb and $^{210}$Po
in OFC using this method and found their contamination for the first time.
Also we observed that the $^{210}$Pb contamination decreases only after 
electrolysis process among copper refinement steps is applied.
We describe details of the method in Section \ref{Sec:Methodology} and
the result of measurements in Section \ref{Sec:Measurement}.

\section{Methodology}
\label{Sec:Methodology}

\subsection{Measurement of $^{210}$Po under low background condition}
\label{Sec:Methodology_measure}
\begin{figure}[htbp]
  \begin{center}
    \includegraphics[width=50mm, angle=270]{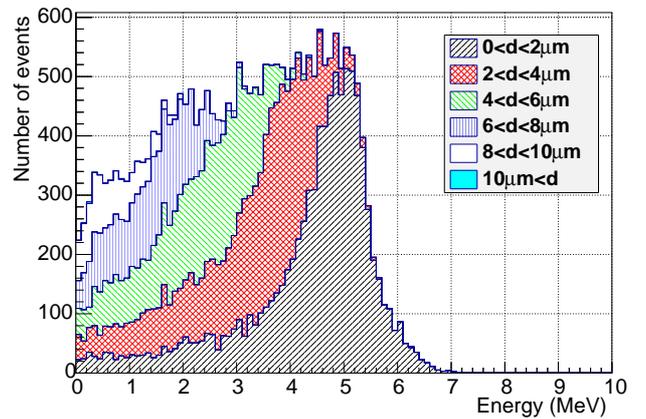}
  \end{center}
  \caption{Energy distribution of $^{210}$Po alpha-ray MC simulation produced in a bulk copper. Stacked histograms are differentiated by the distance ($d$) between alpha-ray generated point and copper surface. Energy resolution of the alpha counter shown in the next subsection is taken into account.}
  \label{fig:Procedure:bulksim}
\end{figure}

A low-background alpha particle counter is commonly used to measure 
the surface RI contamination.
But it is also possible to measure bulk radioactive contamination
\cite{bibBulkPo}.
In the alpha particle counter, energy deposition of an alpha ray emerging from
the material surface is measured.
Fig. \ref{fig:Procedure:bulksim} shows the energy distribution of
5.30 MeV alpha rays generated uniformly in copper bulk obtained from
a Geant4 \cite{bibGeant} Monte Carlo (MC) simulation.
The distribution is continuous because alpha rays emerging from the
bulk lose part of their energy before reaching the surface.
The contribution of bulk $^{210}$Po alpha ray mostly comes from
within 10 $\mu$m from the surface of the copper sample.

To measure bulk contamination, the following conditions are important:
(C1) Background from the alpha particle counter itself should be small.
(C2) Radioactive contamination on the sample surface should be minimized.
(C3) The surface roughness of samples should be much smaller than the range
of alpha rays in copper.
(C4) alpha rays from radioactive nuclei other than $^{210}$Po should be negligible.
To keep these conditions, we performed the following procedures.

In order to satisfy the condition (C1), we used the alpha particle counter,
an Ultra-Lo1800 made by XIA LLC \cite{bibUltraLo}. 
The Ultra-Lo1800 measures the induced charge from ions and electrons
created by an alpha ray.
Background from the outside is vetoed by a surrounding veto counter. 
The measured area of samples should be 30 cm diameter circle or 42.5 cm square.
All the measurements in this paper were performed using the 30 cm diameter
circle mode.
The alpha particle counter was installed underground at Kamioka
Satellite of the Kavli Institute for the Physics and Mathematics of
the Universe, the University of Tokyo, so that the background generated
by cosmogenics is negligible.
 Argon gas is fed from a liquid argon bottle.
In order to reduce the background from emanating $^{222}$Rn and $^{220}$Rn,
the gas line connections are all made from electro-polished (EP)
stainless steel.
Most of the background comes from $^{222}$Rn and $^{220}$Rn emanating
from detector materials or the upstream piping and a liquid argon bottle. 
After the sample is set in the alpha particle counter, argon gas 
is purged for 90 minutes to lower the humidity below $<$50 ppm and also 
to flush out the $^{222}$Rn and $^{220}$Rn.
We do not use the data of the first day, because it still contains $^{222}$Rn
and $^{220}$Rn backgrounds and also the detector conditions such as
humidity become stable after that.
Moreover, we do not use the data within one day before the liquid argon 
bottle runs empty because the energy scale of the detector changes
as the bottle empties.

To satisfy the condition (C2), the alpha particle counter was installed
in a class 1000 clean room in order to minimize dust contamination
during sample exchange. 
We also kept the samples in an ethylene vinyl alcohol copolymer bag
not to accumulate radon daughters on the sample surface during storage
and we assume that any residual accumulation is negligible.
To minimize the residual surface contamination \cite{bibEP}
and to make the surface flat, all copper samples were electro-polished.
By the EP process, the surface becomes smooth with an average roughness
of much less than 1 $\mu$m, that we verified with a laser microscope. 
Therefore the surface roughness is much smaller than alpha ray range in copper,
10 $\mu$m.
So the resulting roughness doesn't affect the alpha counter measurement,
satisfying condition (C3).

In order to confirm the condition (C4),
we measured $^{238}$U and $^{232}$Th bulk contamination in the copper samples
(coarse copper, bare copper, OFC, 6N copper samples shown in Section
\ref{Sec:Measurement}) by GD-MS.
Including the coarse copper which is the lowest purity sample, we found that
all the samples are less than detection limits of the GD-MS, 100 ppt,
which corresponds to 1.2 mBq/kg and 0.4 mBq/kg in $^{238}$U and $^{232}$Th,
respectively. 
The values are the same order of the alpha particle counter background so that
the contributions from $^{238}$U and $^{232}$Th in the copper samples
are negligible.
Also we checked $^{226}$Ra bulk contamination of the coarse copper sample
by HPGe.
No significant signal was found and the upper limit was obtained to be
3.0 mBq/kg. The $^{226}$Ra contribution is negligible.
\vspace{1cm}

\subsection{Data analysis}
\label{Sec:Methodology_analysis}
The alpha counter provides energy ($E$), $t_{0}$, and $t_{\mu}$.
Here, $t_{0}$ is the time when the signal starts to rise and $t_{\mu}$ is
the time when the signal reaches its maximum height.
Therefore, $t_{\mu}$-$t_{0}$ is a proxy for the time electrons drift
from the sample surface to the electrode.
Details are shown in \cite{bibUltraLo}.

\begin{figure}[htbp]
  \begin{center}
    \includegraphics[width=55mm, angle=270]{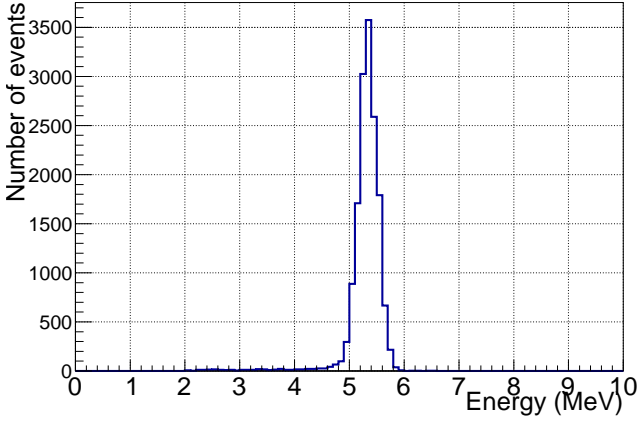}
  \end{center}
  \caption{Energy distribution of the calibration sample. $^{210}$Pb was accumulated on the sample surface. The 5.30 MeV $^{210}$Po alpha-ray peak is clearly seen.}
  \label{fig:Analysis:surface}
\end{figure}

\begin{figure}[htbp]
  \begin{center}
    \includegraphics[width=60mm, angle=270]{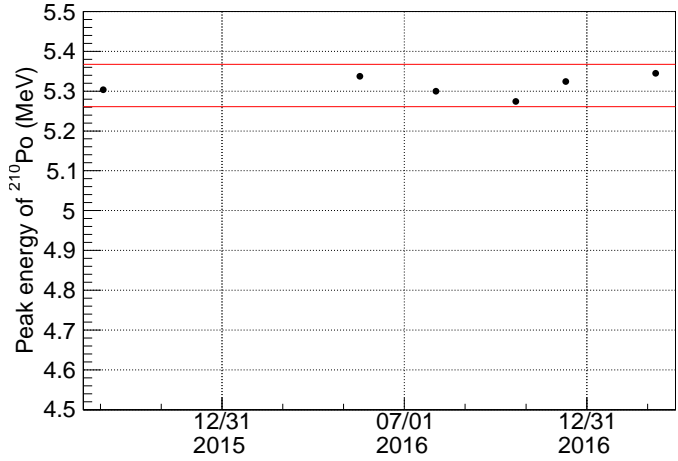}
  \end{center}
  \caption{Time variation of the peak energy of $^{210}$Po alpha ray.
    Red lines delineate $\pm$1\% from the average. The energy scale is 
    stable within 1\%.}
  \label{fig:Analysis:timevariation_gain}
\end{figure}

Energy calibration data are routinely taken using 5.30 MeV
alpha ray from the $^{210}$Po decay of a surface alpha-ray source (10 cm
by 10 cm square copper plate) which is located at the center of the sample tray.
Fig. \ref{fig:Analysis:surface} shows an example of the energy distribution
of 5.30 MeV alpha ray from the $^{210}$Po decay.
Using the data taken just before the sample measurements, the energy is
corrected by fitting the peak.
The energy resolution is estimated to be 4.7\% (1$\sigma$) at 5.30 MeV.
In order to check the stability of the energy, the $^{210}$Po 5.30 MeV peak 
is continuously monitored.
The observed stability is within 1\% as shown in Fig.
\ref{fig:Analysis:timevariation_gain}. 
\vspace{1cm}

Data selection criteria are: (S1) An event is registered, (S2) No veto signal
is observed, (S3) 60 $\mu$s$<$($t_{\mu}$-$t_{0}$), and (S4) 2.5$<E<$4.8 MeV.

\begin{figure}[htbp]
  \begin{center}
    \includegraphics[width=78mm, angle=270]{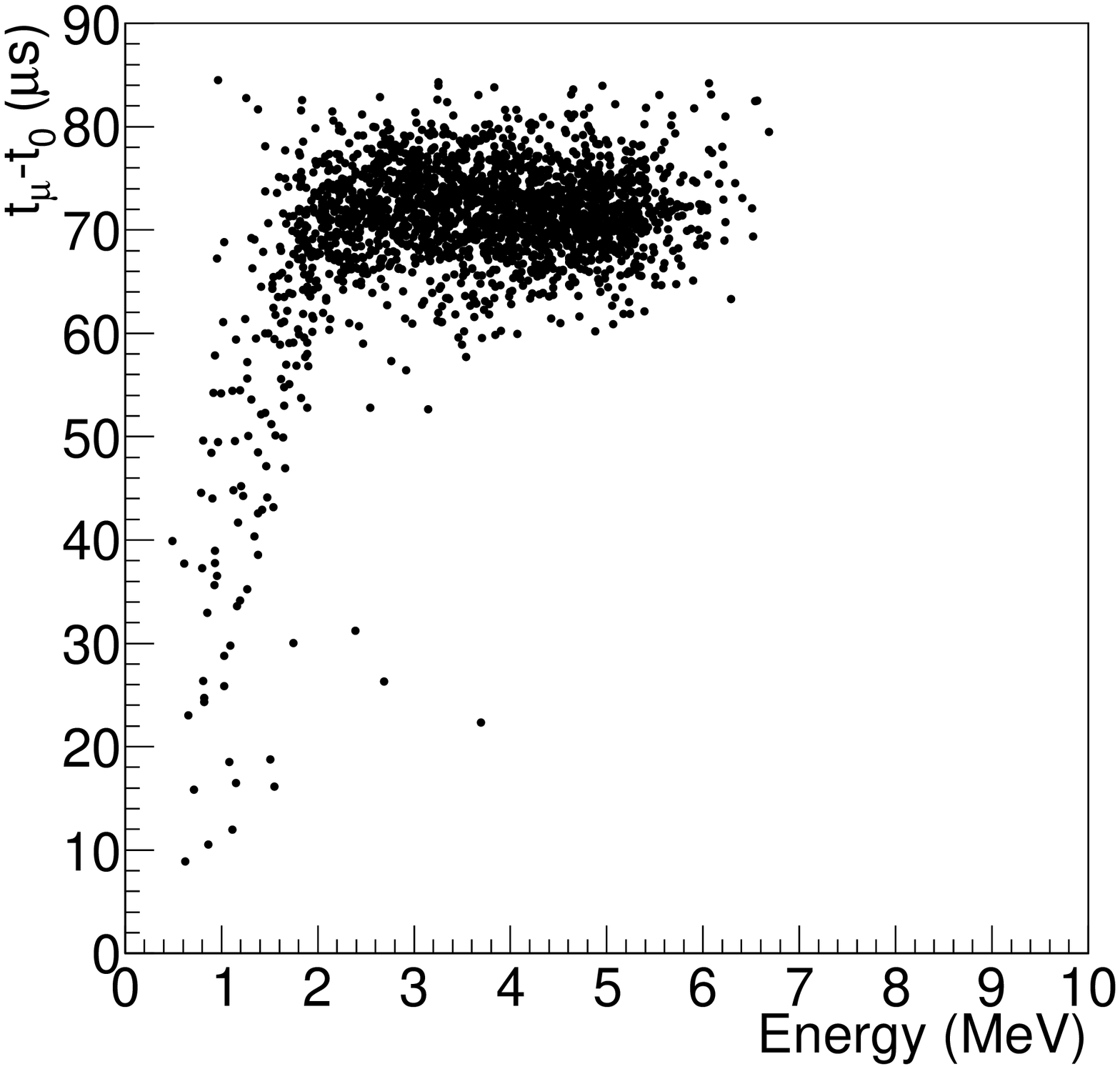}
  \end{center}
  \vspace{0.1cm}
  \begin{center}
    \includegraphics[width=55mm, angle=270]{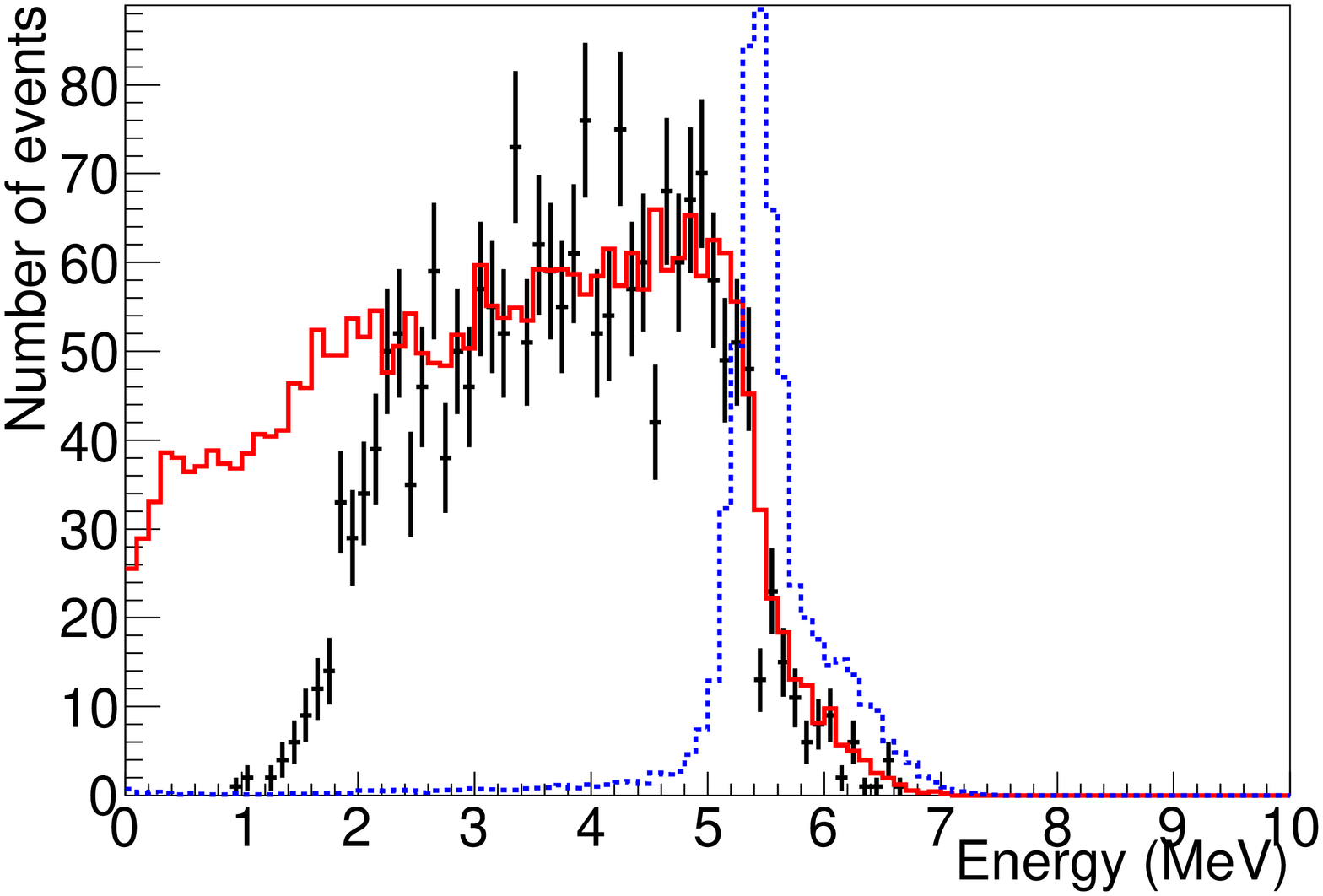}
  \end{center}
  \caption{The upper figure shows Energy vs. $t_{\mu}$-$t_0$ in the coarse copper sample before the $t_{\mu}$-$t_0$ cut. Lower figure overlays the energy distribution of the coarse copper sample after the $t_{\mu}$-$t_0$ cut with the MC simulation of bulk and surface $^{210}$Po. Black points show data with statistical error. The solid and dashed lines show the bulk and surface $^{210}$Po simulation, respectively. Bulk simulation is normalized by number of events in 2.5$<E<$4.8 MeV. Because threshold of the data acquisition is not implemented in the MC, the data at energies $<$2.5MeV doesn't match the MC.}
  \label{fig:Analysis:comparison}
\end{figure}

Fig. \ref{fig:Analysis:comparison} compares data from a 
coarse copper sample provided by Mitsubishi Materials Corporation (MMC)
for our study and the respective simulation.
Coarse copper  is the quality of copper before the electrolysis process
and its purity is about 99\%.
Alpha rays from the bulk have a continuous energy distribution.
As shown in Fig. \ref{fig:Analysis:comparison}, the alpha rays
from the sample surface are clustered around $t_{\mu}$-$t_{0}$ = 70 $\mu$s.
Criterion (S3) selects events happening at the sample surface, not
in the gas surrounding sample.

We only use the data where the energy is larger than 2.5 MeV because the
efficiency below this energy is affected by the trigger threshold effects
as evident in Fig. \ref{fig:Analysis:comparison}.
Events around 5.30 MeV may contain residual surface $^{210}$Po alpha-ray
events. 
Therefore events in 2.5$<E<$4.8 MeV are used to estimate 
the $^{210}$Po bulk contamination.
The contribution in $E<$4.8 MeV from surface $^{210}$Po alpha rays is
estimated to be less than 10\%, based on the calibration sample study.
Energy distribution of the surface $^{210}$Po simulation is sharper than
that of the calibration data shown in Fig. \ref{fig:Analysis:comparison}.
It is because the simulation assumed the contamination was evenly
distributed over the entire surface, though the calibration sample
is smaller (10 cm by 10 cm) and located at the center of the sample tray.
Energy resolution at the outer boundary of the sample area is not as
good as that in the central area, that is taken into account in the simulation.

We use the emissivity in 2.5$<E<$4.8 MeV to estimate the $^{210}$Po
contamination in the copper bulk.
Emissivity is the efficiency corrected alpha-ray event rate from 
a unit area.
From the MC simulation, the conversion factor from the emissivity 
in 2.5$<E<$4.8 MeV to $^{210}$Po contamination for copper is obtained as
2.7$\times$10$^{2}$ (Bq/kg)/(alpha/cm$^2$/hr).
\vspace{1cm}

The observed background level in our setup before applying criterion (S4)
is about 10$^{-4}$ alpha/cm$^{2}$/hr.
The remaining background comes from the decay of daughter nuclei of
$^{220}$Rn and $^{222}$Rn (down to $^{210}$Pb) to the surface of the samples.
The energies of the dominant alpha rays from these daughters are 6.78 MeV from
$^{216}$Po, 6.05 MeV from $^{212}$Bi, 8.79 MeV from $^{212}$Po, 6.00 MeV
from $^{218}$Po, and 7.69 MeV from $^{214}$Po.
The lowest energy of alpha ray among these daughter nuclei is
6.00 MeV from the $^{218}$Po decay.
Thus, most of background can be removed by selecting
low-energy events and the background for the bulk $^{210}$Po
measurement is reduced to be $<$10$^{-5}$ alpha/cm$^2$/hr.

The remaining background from the alpha particle counter was estimated by
measuring a clean silicon wafer.
Simply measuring the emissivity of an empty stainless steel sample tray is
not relevant as any sample put on the tray prevents that background
from reaching the alpha particle counter.
A silicon wafer is one of the cleanest among our measured sample.
Assuming that it has no bulk contribution in itself we use this measurement
to extract a maximal contribution from the detector to the 2.5$<E<$4.8 MeV
region.
The measured emissivity of the silicon wafer in 2.5$<E<$4.8 MeV is
(5.6$\pm$5.6)$\times$10$^{-6}$ alpha/cm$^2$/hr, which would correspond
to a 1.5 mBq/kg $^{210}$Po contamination using the MC derived copper
bulk conversion factor.

\begin{figure}[htbp]
  \begin{center}
    \includegraphics[width=55mm, angle=270]{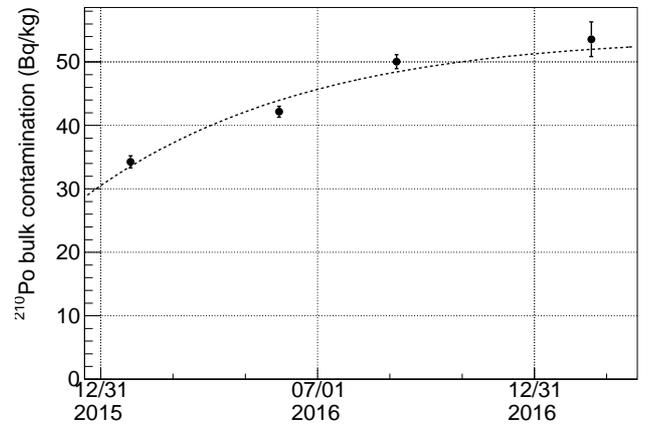}
  \end{center}
  \caption{Time evolution of event rate of $^{210}$Po in coarse copper bulk. Line shows a fitted result using Eq. \ref{Equ:decay1} and \ref{Equ:decay2}.}
  \label{fig:Analysis:timevariation_coarse}
\end{figure}

To determine the $^{210}$Pb contamination, we measure $^{210}$Po several
times over different periods because the $^{210}$Pb and $^{210}$Po decay
equilibrium may have been previously broken.
The evolution of the $^{210}$Pb and $^{210}$Po contaminations is described
as the following equations:
\begin{equation}
\label{Equ:decay1}
  N_0(t) = N_0(0)\exp(-t/\tau_0)
\end{equation}
\begin{eqnarray}
\label{Equ:decay2}
  N_1(t) = N_0(0)\frac{\frac{1}{\tau_0}}{\frac{1}{\tau_1}-\frac{1}{\tau_0}}(\exp(-t/\tau_0)-\exp(-t/\tau_1)) \nonumber\\
+N_1(0)\exp(-t/\tau_1),
\end{eqnarray}
where $\tau_0$ and $\tau_1$ are the life times of $^{210}$Pb and 
$^{210}$Po, and they are 32.17 years and 0.55 years, respectively.
$N_0(t)$ and $N_1(t)$ are the $^{210}$Pb and $^{210}$Po contaminations
at time $t$, respectively.
$N_1(t)$ is obtained from the emissivity measurement after background
subtraction.
$t$=0 is the time when the copper samples were delivered.
By fitting the data with the equations above, $^{210}$Pb and $^{210}$Po
contaminations are obtained.
Fig. \ref{fig:Analysis:timevariation_coarse} shows the time evolution
to the equilibrium of the $^{210}$Po contamination in the coarse copper
bulk together with the resulting fit.
The $^{210}$Pb contamination in coarse copper bulk is estimated to be 
57$\pm$1 Bq/kg. 
To check the validity of the conversion factor, the $^{210}$Pb contamination
in coarse copper is investigated with HPGe.
The measured contamination is 55.6$\pm$1.5 (stat.)+16.7-5.6 (sys.) Bq/kg,
which is consistent with our alpha particle counter measurement.
The dominant systematic uncertainty of the conversion factor
comes from the uncertainty of the HPGe measurement and is estimated to be
+30\% -10\%.
\vspace{1cm}

\section{Copper sample measurements}
\label{Sec:Measurement}
Based on the methodology detailed in the previous section, we measure the
$^{210}$Pb and ${^{210}}$Po contaminations in various copper samples.

\subsection{Oxygen free copper samples}
\begin{figure}[htbp]
  \begin{center}
    \includegraphics[width=55mm, angle=270]{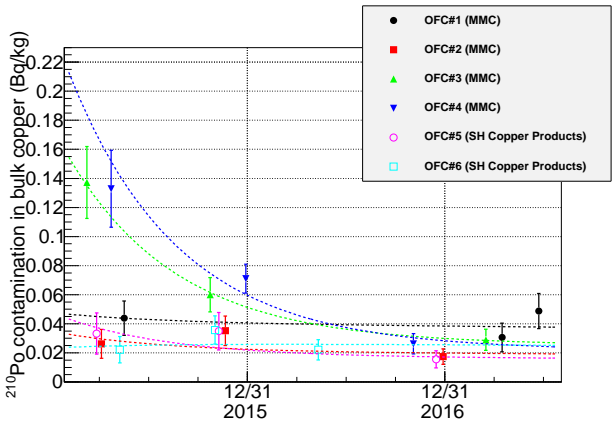}
  \end{center}
  \caption{Time evolution of $^{210}$Po bulk contamination in our OFC samples. Error bars show the statistical error. Sample \#1-4 were made by MMC. Sample \#5 and 6 were made by SH copper products. The samples were delivered on Jan. 31st, 2015.}
  \label{fig:Analysis:timevariation_OFC_bulk}
\end{figure}

We investigate different OFC (Japanese Industrial Standards (JIS)
H2123 C1020) batches as well as OFC samples made by different companies.
We purchased plates that were processed by rolling.
All plates were electro-polished taking off about 20 $\mu$m of the
material surface.
Fig. \ref{fig:Analysis:timevariation_OFC_bulk} shows the time evolution
of the $^{210}$Po contaminations in the bulk for the measured samples.
Four samples were made by MMC and the other two by another company,
SH copper products.
As a result of the fitting, the $^{210}$Pb contaminations in all
the samples are estimated to be within 17-40 mBq/kg, though $^{210}$Po
contaminations vary widely as summarized in Table \ref{tbl:Result}.
This is the first identification of a $^{210}$Pb contamination in OFC.
Although OFC is known as a clean material, we found that our samples contain
a few tens of mBq/kg of $^{210}$Pb.
The variation of the $^{210}$Pb contaminations among the OFC samples is small.
We also investigated the highest grade of OFC, class 1 according to the
American Society for Testing and Materials (ASTM) B170 C10100 standard.
It was made by SH copper products and the $^{210}$Pb contamination we
measured was 36$\pm$13 mBq/kg.
Thus we do not see a significant difference on the $^{210}$Pb contamination
between the different OFC grades.
\vspace{1cm}

\subsection{Copper samples from different steps in the production process}
\label{Sec:prod}
\begin{figure}[htbp]
  \begin{center}
    \includegraphics[width=80mm, angle=0]{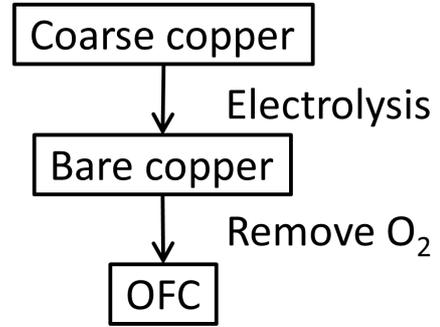}
  \end{center}
  \caption{Copper sample production summary.}
  \label{fig:Analysis:Production}
\end{figure}

\begin{figure}[htbp]
  \begin{center}
    \includegraphics[width=55mm, angle=270]{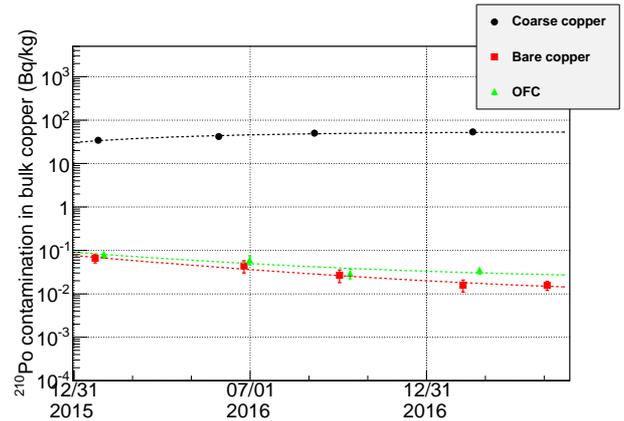}
  \end{center}
  \caption{Time evolution of the $^{210}$Po contamination in coarse copper, bare copper, and OFC. Error bars show the statistical error. The samples were delivered on Oct. 5th, 2015.}
  \label{fig:Analysis:timevariation_bulk}
\end{figure}

In order to investigate the origin of the $^{210}$Pb contamination in
OFC, we measured copper samples under refining, namely, coarse copper,
bare copper, and OFC (JIS H2123 C1020). 
These were all made and provided by MMC for this study.
The result of the coarse copper measurement was already shown in
Section \ref{Sec:Methodology}.
Bare copper is made from coarse copper by electrolysis.
OFC is then made from bare copper by removing the oxygen contamination
as summarized in Fig. \ref{fig:Analysis:Production}.
The samples were all formed by cutting and milling, but not by rolling
in order to minimize possible $^{210}$Pb accumulation from roller.
To remove the contribution from surface activity,
we removed about 50 $\mu$m surface materials by EP with fresh
electrolysis solution in a clean room.
Each sample was measured more than three times to investigate time evolution
of $^{210}$Pb and $^{210}$Po decays.
Fig. \ref{fig:Analysis:timevariation_bulk} shows the time evolution of
$^{210}$Po contaminations of the samples.
The fit results including other samples are summarized in
Table \ref{tbl:Result}.
Coarse copper has the highest $^{210}$Pb contamination. 
The $^{210}$Pb contamination of the samples that underwent
the electrolysis process, bare copper and OFC, are more than three
orders of magnitude lower than that of coarse copper.
The equilibrium between $^{210}$Pb and $^{210}$Po decays is largely broken.
In the coarse copper sample, the $^{210}$Pb contamination is higher than
$^{210}$Po contamination.
On the other hand, bare copper and OFC have a higher $^{210}$Po contamination
than $^{210}$Pb contamination.
These results indicate most of the $^{210}$Pb is removed in the electrolysis
process but the reduction ratio for the daughter isotopes,
$^{210}$Bi or $^{210}$Po is smaller than that of $^{210}$Pb. 
The $^{210}$Pb contamination in OFC that was not rolled is consistent
with that of the rolled OFC samples shown in Fig. \ref{fig:Analysis:timevariation_OFC_bulk}.
\vspace{1cm}

\begin{table*}[tbp]
  \begin{center}
    \caption{Summary of measurements. The contamination when the samples were delivered are shown. The errors show the statistical error. The systematic uncertainty is +30\% -10\%.}
    \begin{tabular}{ccc}
      \hline \hline
      Sample & $^{210}$Pb contamination &  $^{210}$Po contamination\\
      & (mBq/kg) &  (mBq/kg)\\
      \hline
      OFC\#1 (C1020) (MMC) & 40$\pm$8 & 47$\pm$21\\
      OFC\#2 (C1020) (MMC) & 20$\pm$6 & 33$\pm$14\\
      OFC\#3 (C1020) (MMC) & 27$\pm$7 & (1.6$\pm$0.3)$\times$10$^2$\\
      OFC\#4 (C1020) (MMC) & 23$\pm$8 & (2.2$\pm$0.4)$\times$10$^2$\\
      OFC\#5 (C1020) (SH copper products) & 17$\pm$6 & 44$\pm$18\\
      OFC\#6 (C1020) (SH copper products) & 27$\pm$8 & 24$\pm$17\\
      OFC (class1) (SH copper products) & 36$\pm$13 & 38$\pm$3\\
      \hline
      Coarse copper (MMC) & (57$\pm$1)$\times$10$^3$ & (16$\pm$2)$\times$10$^3$\\
      Bare copper (MMC) & 8.4$\pm$4.0 & (1.1$\pm$0.2)$\times$10$^2$\\
      OFC (MMC) & 23$\pm$8 & (1.3$\pm$0.3)$\times$10$^2$\\
      \hline
      6N copper (MMC) & $<$4.1 & $<$4.8\\
      Electroformed copper (Asahi-Kinzoku) & $<$5.3 & $<$18\\
      \hline \hline
    \end{tabular}
    \label{tbl:Result}
  \end{center}
\end{table*}    

\subsection{Very low $^{210}$Pb contamination copper}
\label{Sec:lowpb}
To find the copper least contaminated with $^{210}$Pb, we investigated
6N ($>$99.9999\% purity) copper made by MMC.
6N copper is made from bare copper.
The main process to make 6N copper is an additional electrolysis in a clean
environment. 
The number of $^{210}$Po events in the 6N copper sample is only one or two
events over two weeks measurement. 
This event rate from the 6N copper sample is consistent with the background
observed in the silicon wafer samples. 
This implies that the copper is as pure as the silicon wafer or that
both materials purity exceed the sensitivity capabilities of the instruments.
Therefore we can only derive an upper limit for the $^{210}$Pb contamination
in 6N copper.
The $^{210}$Pb and $^{210}$Po contamination limits at 90\% confidence level
 (CL) are estimated to be $<$4.1 mBq/kg and $<$4.8 mBq/kg, respectively.
The $^{210}$Pb contamination of 6N copper is much smaller than 
those of OFC. 
According to the GD-MS measurement, one of the stable lead isotope,
$^{208}$Pb contaminations in coarse copper, bare copper, OFC,
and 6N copper are 2200 ppm, 0.19 ppm, 0.98 ppm, and 0.002 ppm, respectively.
The $^{208}$Pb contamination ratio of coarse copper, bare copper, and OFC are
similar to that of our $^{210}$Pb measurements.
The $^{210}$Pb contamination in 6N copper is expected to be 0.05 mBq/kg
if we assume that the ratio between $^{208}$Pb and $^{210}$Pb is common among
these copper samples.
To directly measure such a low level contamination, further background
reduction and longer duration measurements are required.

Another very low $^{210}$Pb contamination sample was electroformed copper.
The electroformed copper we measured was accumulated on a stainless
steel base about 500 $\mu$m in thickness.
The input material was copper phosphate.
After machining the accumulated copper to make flat surface, EP is applied
to remove about 50 $\mu$m of materials.
The observed number of alpha-ray events was consistent with the background
expectation. 
The $^{210}$Pb contamination limit at 90\% CL in the electroformed
copper bulk is estimated to be $<$5.3 mBq/kg.

\section{Summary and discussion}
We established a new method to measure the $^{210}$Pb and $^{210}$Po
contaminations in copper bulk by distinguishing bulk originating alpha rays
from those of surface. 
This is the first measurement of $^{210}$Pb contamination
in copper bulk using a low-background alpha particle counter.
Due to the achievement of very low background in the alpha counter,
the sensitivity to measure this $^{210}$Pb and $^{210}$Po contaminations
reaches a few mBq/kg.
We found that OFC contains a small, but non-zero, 17-40 mBq/kg contaminations
of $^{210}$Pb.
This $^{210}$Pb is interpreted as a small residual from the electrolysis
process based on the investigation of the coppers in each production step. 
The 6N and the electroformed coppers contaminate $^{210}$Pb at least one
order lower than OFC.
The method we established here to measure the copper bulk contamination could
apply to other materials.
Because radon emanates continuously into the air and decay products are
accumulated on the surface of the material, $^{210}$Pb can get mixed into
the material in the production process.
$^{210}$Pb contamination needs to be investigated independently
from that of $^{238}$U or $^{226}$Ra.
To check the $^{210}$Pb bulk contamination is important for dark matter
and neutrinoless double beta decay experiments.
Our method can improve material screening and preparation for such low
background experiments.

	\section*{Acknowledgments}
	We gratefully acknowledge the cooperation of Mitsubishi Materials Corporation. They kindly provide us the copper samples in each copper production process, that helps the understanding of the origin of $^{210}$Pb in OFC.
	We also gratefully acknowledge the cooperation of Kamioka Mining  and  Smelting  Company.
   This work was supported by the Japanese Ministry of Education, Culture, Sports, Science and Technology, Grant-in-Aid for Scientific Research, ICRR Joint-Usage, JSPS KAKENHI Grant Number, 26104004.



\section*{References}
	\begin{thebibliography}{99}
		\bibitem{bibXMASSDetector} K.~Abe {\it et al.}, Nucl.\ Instrum.\ Meth.\ A {\bf 716} (2013) 78.
		\bibitem{bibCUOREDetector} C.~Alduino {\it et al.},  Journal of Instrumentation 11, P07009 (2016).
        \bibitem{bibBulkPo} G.~Zuzel {\it et al.}, Applied Radiation and Isotopes {\bf 126} (2017) 165-167.
        \bibitem{bibGeant} S. ~Agostinelli {\it et al.}, Nucl.\ Instrum.\ Meth.\ A {\bf 506} (2003) 250.
        \bibitem{bibUltraLo} W.~K.~Warburton, J.~Wahl, and M.~Momayezi, Ultra-low Background Gas-filled Alpha Counter, U.S. Patent 6 732 059, May 4, 2004.
          W.~K.~Warburton {\it et al.}, IEEE Nuclear Sci. Symp. Conf. Rec., Oct. 16-22, 2004, vol. 1, pp. 577-581, Paper N16-80.
          M.~Z.~Nakib {\it et al.}, AIP Proc. 1549, 78-81 (2013).
          B.~D.~McNally {\it et al.}, Nucl.\ Instrum.\ Meth.\ A {\bf 750} (2014) 96-102.
        \bibitem{bibEP} G.~Zuzel {\it et al.}, Nucl.\ Instrum.\ Meth.\ A {\bf 676} (2012) 140-148.
	\end{thebibliography}
	\end{document}